\documentclass[twocolumn]{bmcart}%

\usepackage[utf8]{inputenc} %

\def\includegraphics{}

\startlocaldefs
\endlocaldefs
\usepackage{colortbl}
\usepackage{color}
\usepackage{multirow}
\usepackage{siunitx}
\usepackage{graphicx}
\DeclareSIUnit\percentpoint{pp}
\DeclareSIUnit\photoelectron{p.e.}
\DeclareSIUnit\bit{bit}
\definecolor{color0}{rgb}{1,1,0}

\usepackage{type1cm}
\usepackage{eso-pic}

	\makeatletter
	\AddToShipoutPicture{%
		\setlength{\@tempdimb}{.5\paperwidth}%
		\setlength{\@tempdimc}{.5\paperheight}%
		\setlength{\unitlength}{1pt}%
		\put(\strip@pt\@tempdimb,\strip@pt\@tempdimc){%
			\makebox(0,0){\rotatebox{45}{\textcolor[gray]{0.75}%
					{\fontsize{3cm}{3cm}\selectfont{authors' manuscript}}}}%
		}%
	}
\RequirePackage[normalem]{ulem} %
\RequirePackage{color}\definecolor{RED}{rgb}{1,0,0}\definecolor{BLUE}{rgb}{0,0,1} %
\providecommand{\DIFaddbegin}{} %
\providecommand{\DIFaddend}{} %
\providecommand{\DIFdelbegin}{} %
\providecommand{\DIFdelend}{} %

\begin{document}

\begin{frontmatter}

\begin{fmbox}
\dochead{Original Research}

\DIFaddbegin 

\DIFaddend \title{Evaluation Of The PETsys TOFPET2 ASIC In Multi-Channel Coincidence Experiments}

\DIFdelbegin %
\DIFdelend %

\DIFdelbegin %
\DIFdelend \DIFaddbegin \author[
   addressref={aff1},                   %
]{\inits{VN}\fnm{Vanessa} \snm{Nadig}}

\author[
addressref={aff1,aff2},                   %
]{\DIFaddend \inits{DS}\fnm{David} \snm{Schug}}

\DIFdelbegin %
\DIFdelend \DIFaddbegin \author[
addressref={aff1,aff2},                   %
]{\DIFaddend \inits{BW}\fnm{Bjoern} \snm{Weissler}}

\DIFdelbegin %
\DIFdelend \DIFaddbegin \author[
addressref={aff1,aff2,aff3,aff4},                   %
corref={aff1},                       %
email={volkmar.schulz@pmi.rwth-aachen.de}   %
]{\inits{VS}\fnm{Volkmar} \snm{Schulz}}
\DIFaddend

\DIFdelbegin %
\DIFdelend %

\DIFdelbegin %
\DIFdelend \DIFaddbegin \address[id=aff1]{%
  \orgname{Department of Physics of Molecular Imaging Systems, Experimental Molecular Imaging, RWTH Aachen University}, %
  \street{Pauwelsstrasse 17},                     %
  \postcode{52074},                                %
  \city{Aachen},                              %
  \cny{GER}                                    %
}
\DIFaddend 

\DIFdelbegin %
\DIFdelend \DIFaddbegin \address[id=aff2]{%
	\orgname{Hyperion Hybrid Imaging Systems GmbH}, %
	\street{Pauwelsstrasse 19},                     %
	\postcode{52074},                         %
	\city{Aachen},                              %
	\cny{GER}                                    %
}
\DIFaddend 

\DIFdelbegin %
\DIFdelend \DIFaddbegin \address[id=aff3]{%
	\orgname{III. Physikalisches Institut B, RWTH Aachen University}, %
	\street{Otto-Blumenthal-Straße},                     %
	\postcode{52074},                             %
	\city{Aachen},                              %
	\cny{GER}                                    %
}

\address[id=aff4]{%
	\orgname{Fraunhofer Institute for Digital Medicine MEVIS}, %
	\street{Forckenbeckstrasse 55},                     %
	\postcode{52074},                           %
	\city{Aachen},                              %
	\cny{GER}                                    %
}

\DIFaddend 

\begin{artnotes}
\end{artnotes}

\DIFaddbegin

\DIFaddend \textcolor{red}{Authors' revised manuscript submitted to EJNMMI Physics February 2021}

\begin{abstractbox}

\begin{abstract} %

\parttitle{Background} 
Aiming to measure the difference in arrival times of two coincident $\gamma$-photons with an accuracy in the order of \SI{200}{\pico\second}, time-of-flight positron emission tomography systems commonly employ silicon photomultipliers (SiPMs) and high-resolution digitization electronics, application specific integrated circuits (ASICs).
This work evaluates the performance of the TOFPET2 ASIC, released by PETsys Electronics S.A. in 2017, dependent on its configuration parameters in multi-channel coincidence measurements.

\parttitle{Methods}
SiPM arrays fabricated by different vendors (KETEK, SensL, Hamamatsu, Broadcom) were tested in combination with the ASIC.
Scintillator arrays featuring different reflector designs and different configurations of the TOFPET2 ASIC software parameters were evaluated.
The benchtop setup used is provided with the TOFPET2 ASIC evaluation kit by PETsys Electronics S.A..

\parttitle{Results}
Compared to existing studies featuring the TOFPET2 ASIC, multi-channel performance results dependent on a larger set of ASIC configuration parameters were obtained that have not been reported to this extend so far.   
The ASIC shows promising CRTs down to \SI{219.9}{\pico\second} in combination with two Hamamatsu S14161-3050-HS-08 SiPM arrays (128 channels read out, energy resolution \SI{13.08}{\percent}) and \SI{216.1}{\pico\second} in combination with two Broadcom AFBR-S4N44P643S SiPM arrays (32 channels read out, energy resolution \SI{9.46}{\percent}).
The length of the trigger delay  of the dark count suppression scheme has an impact on the ASIC performance and can be configured to further improve the coincidence resolution time.
The integrator gain configuration has been investigated and allows an absolute improvement of the energy resolution by up to \SI{1}{\percent} at the cost of the linearity of the energy spectrum.

\parttitle{Conclusion}
Measuring up to the time-of-flight performance of state-of-the-art positron emission tomography (ToF-PET) systems while providing a uniform and stable readout for multiple channels at the same time, the TOFPET2 ASIC is treated as promising candidate for the integration in future ToF-PET systems.

\end{abstract}

\DIFaddbegin 

\DIFaddend 

\end{abstractbox}
\DIFdelbegin %
\DIFdelend %
\DIFaddbegin \end{fmbox}%
\DIFaddend 

\end{frontmatter}

\DIFaddbegin

\DIFaddend \section{Background}

A functional imaging modality widely used in the diagnosis and staging of cancer as well as in cardiology or neurology is positron emission tomography (PET) 
\cite{bailey2005pet,phelps2006pet,vaquero2015PETreview}.
After an injected tracer undergoing a $\beta^{+}$-decay, the emitted positron annihilates with an electron in the surrounding tissue, resulting in the back-to-back release of two $\gamma$-photons. %
The $\gamma$-photons are detected by two opposing elements of a ring-shaped detector each consisting of a scintillator array coupled to a photo-detector array.
Via a scintillation process, the $\gamma$-photons are stopped and converted in to optical photons, which then reach the employed photo-detector.\\
In time-of-flight positron emission tomography (ToF-PET), the difference in arrival times of the two detected $\gamma$-photons can be resolved, which results in a more precise localization of the annihilation event along the line of response (LOR) connecting the two points of detection \cite{surti2016advances,vandenberghe2016RecentDevelopments}.
The coincidence resolution time (CRT) assesses the capability of a PET system to resolve this ToF information.
Studies have shown that incorporating ToF information into the image reconstruction process increases the signal-to-noise ratio (SNR) of a PET image and therefore improves the image quality \cite{conti2011focusToF,gundacker2014ToFGain,surti2015ToFGain}.
For a precise measurement of the ToF information, CRTs in the order of few hundred picoseconds are required.
A low energy resolution in the order of \SI{10}{\percent} around 511 keV is of advantage to filter true coincidences from scattered events.\\
State-of-the-art clinical PET systems reach CRTs ranging from \SI{214}{\pico\second} to \SI{500}{\pico\second}. %
The achieved energy resolutions lay between \SI{9}{\percent} and \SI{12}{\percent} 
\cite{vandenberghe2016RecentDevelopments,Spencer2020}.
For prototype PET systems, CRTs between \SI{200}{\pico\second} to \SI{450}{\pico\second} and energy resolutions between \SI{11}{\percent} and \SI{12}{\percent} are reported 
\cite{vandenberghe2016RecentDevelopments,schug2015tofring}.
For both cases, future systems aim to reach CRTs below \SI{200}{\pico\second} \cite{bugalho2017ExperimentalResultsTOFPET2,chen2014DedicatedReadoutASIC,rolo2012TOFPETASIC}.
On benchtop level, this limit has already been excelled by various setups reaching CRTs below \SI{100}{\pico\second} \cite{nemallapudi2015sub100ps,sarasola2017FlexToTvsNINO,gundacker2019CTR58ps}.\\
Scintillators used in PET systems require a high stopping power, a short scintillation decay time, and high photon statistics \cite{lecoq2016scintillators}.
Reflective foils or mixtures of glues and powders can be used to optically segment an array of scintillators.
Here, various scintillator topologies can be considered, namely one-to-one coupling between single scintillator needles and photo-sensor channels, multiple scintillator needles on one photo-sensor channel as well as monolithic scintillator blocks or slabs on an array of photo-sensor channels. The latter ones allow for depth-of-interaction (DOI) positioning \cite{borghi2016Monoliths,mueller2018Monotliths,mueller2018GTB,peng2019ComptonPET,grosswege2016MLPositioning,schug2017timewalk}.\\
In common state-of-the-art systems, photo-detectors such as photo-multiplier tubes (PMTs) and analog or digital silicon photomultipliers (SiPMs) and custom-designed readout electronics are employed.
SiPMs became popular due to the high photo detection efficiency (PDE) (up to \SI{50}{\percent}, some up to \SI{65}{\percent}), their high internal gain, their compactness, their fast response time, and their compatibility with magnetic fields, allowing simultaneous PET/MR imaging  \cite{bisogni2016SiPMDevelopment,bisogni2018medicalapplicationsSiPMs,gundacker2019CTR58ps,weissler2014analogPETMR,weissler2015MRtests}.
SiPMs consist of several thousands single-photon avalanche diodes (SPADs) which are connected in parallel and operated in Geiger-mode. %
An incident optical photon hitting a SPAD causes a self-sustaining charge carrier avalanche.
The avalanche effect is used for signal amplification enabling the detection of single optical photons, whereby individual SPAD signals are overlaid to a sum signal forming an SiPM pulse with a steep rising edge \cite{bisogni2016SiPMDevelopment}.
The first prototype system making use of digital SiPMs (dSiPMs) featuring direct binary digitization has been developed in recent years \cite{frach2019dSiPM,haemisch2012DPC,schaart2015dSiPM, hallen2018PETeval, schug2016petperformance}.\\
Typically, application-specific integrated circuits (ASICs) are employed to precisely digitize analog SiPM signals.
This study specifically focuses on the application of ASICs in ToF-PET systems.
Apart from exclusive ToF-PET systems, also hybrid imaging systems combining ToF-PET with other modalities, such as magnetic resonance imaging (MRI), electroencephalography (EEG) or ultrasound imaging (US), feature ASICs as readout and digitization electronics (MADPET4 \cite{omidvari2018MRcompatibilityMADPET4,omidvari2017MADPET4}, TRIMAGE \cite{ahmad2015Triroc,ahmad2016Triroc,sportelli2016TRIMAGEInitialResults}, EndoToFPET-US \cite{chen2014DedicatedReadoutASIC,shen2012STiC}).
ASICs typically feature a comparator with a low threshold to trigger on the rising edge of the SiPM pulse, i.e., on the first optical photons arriving, and a time-to-digital converter (TDC) to assign a timestamp to an event.
The energy of the respective event can then be measured either by signal integration featuring capacitors (qdc-method) or by measuring the time over a specified threshold (tot-method) \cite{chen2014DedicatedReadoutASIC,corsi2009ASICdevelopment,orita2017CurrentToTASIC}.
The Weeroc series \cite{ahmad2015Triroc,ahmad2016Triroc,fleury2017PETIROC2A}, the PETA series \cite{fischer2009PETA,fischer2006PETAseries,piemonte2012PerformancePETA3ASIC,sacco2013PETA4,sacco2015PETA5,schug2017PETA5,weissler2014analogPETMR} and the TOFPET2 series by PETsys Electronics S.A. \cite{PETsysTOFPET2DataSheet,PETsysTOFPET1DataSheet} use such a qdc-method to measure the signal energy.\\
In this work, we evaluate the multi-channel performance of the TOFPET2 ASIC (version 2b), released by PETsys Electronics S.A. in 2017 \cite{PETsysWebPage,schug2018TOFPET2,bugalho2019TOFPET2characterization}.
Existing studies on the TOFPET2 ASIC performance deal with simulations or experimental results for single SiPMs in combination with the ASIC, which greatly differs to the experimental multi-channel results we are providing \cite{bugalho2017ExperimentalResultsTOFPET2,francesco2016TOFPET2,schug2018TOFPET2}. 
Further studies show multi-channel results that are restricted to a very small parameter space or focus on the DOI resolution of a certain scintillator topology obtained with TOFPET2 readout \cite{bugalho2019TOFPET2characterization,li2019DOI,lamprou2020TOFPET2Monoliths,lamprou2018CharacterizationTOFPET}.
They do not report experimental performance results dependent on the TOFPET2 configuration parameters, e.g., on the delay line configuration or the integrator gain, to the extend we do.\\
The goal of this study is to assess whether the TOFPET2 ASIC should be considered as solution for digitizing and processing analog SiPM signals in future (whole-body) ToF-PET systems.
The presented multi-channel studies are necessary to investigate detector-related effects such as crosstalk and light-sharing as well as the channel-spread of the ASIC performance handling data from multiple channels at once, which cannot be assess in single-channel measurements.
The compatibility of the ASIC with different SiPM types in combination with different materials for scintillator segmentation was tested.
In addition, the parameters of the ASIC configuration related to the trigger circuit were varied. 
The impact of the hard- and software configuration on the ASIC performance was quantified.

\section{Materials and Methods}

\DIFaddbegin 

\DIFaddend \subsection{TOFPET2 ASIC} 

\begin{figure}[h!]
	\includegraphics[width=0.45\textwidth]{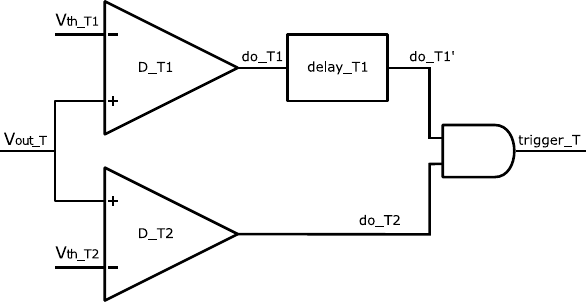}
	\caption{\csentence{Schematic drawing of the TOFPET2 channel trigger circuit.} 
		In the timing branch of the trigger circuit, the output of the first discriminator $\mathsf{D\_T1}$ is delay by a configurable delay element with respect to the output of the second discriminator $\mathsf{D\_T2}$. The design allows the rejection of small noise pulses.}
	\label{fig:schematic_circuit}
\end{figure}

The TOFPET2 ASIC is characterized by its compactness (\SI{14x14}{\mm} chip size including bonding area \cite{PETsysFlyer}), 128 readout channels, low power consumption \cite{PETsysTOFPET2DataSheet,nadig2019TOFPET2PowerConsumption} and a high hit rate of up to \SI{480}{kcps} per channel, which corresponds to a data rate of \SI{640}{\mega \bit/\second} \cite{PETsysTOFPET2DataSheet,francesco2016TOFPET2}. 
The maximum output data rate of the ASIC, i.e., 64 channels, is specified as \SI{2.8}{\giga \bit/\second} \cite{PETsysTOFPET2DataSheet}.\\
Each of the 64 individual channels is multi-buffered by four analog buffers and employs a three-threshold trigger logic with two discriminators $\mathsf{D\_T1}$ and $\mathsf{D\_T2}$ in the timing and one discriminator $\mathsf{D\_E}$ in the energy branch \cite{PETsysTOFPET2DataSheet,TOFPET2EKitHardwareGuide,TOFPET2EKitSoftwareGuide}.
A global- and a channel-specific configuration register allow to change the ASIC configuration.
The ASIC can be operated in a time-over-threshold (tot) or an energy integration (qdc) mode for energy measurement.
The charge-to-digital converter (QDC) used in qdc mode behaves linear for integration charges up to \SI{1500}{\pico\coulomb} \cite{francesco2016TOFPET2}.
The TDC has a resolution of \SI{30}{\pico\second}.
The chip runs with a clock cycle of \SI{200}{\mega\hertz}.\\
Incoming SiPM signals are amplified by a trans-impedance ampliflier in the timing branch (nominal gain \SI{3000}{\ohm}) and a transimpedance ampliflier energy branch (nominal gain \SI{300}{\ohm}), respectively \cite{PETsysTOFPET2DataSheet}.
As determined in an oscilloscope measurement, the pre-amplified SiPM pulses in the timing branch have a signal height of about \SI{300}{\milli\volt}.
The thresholds of the three discriminators can be adjusted via the three dimensionless parameters $\mathsf{vth\_t1}$, $\mathsf{vth\_t2}$, and $\mathsf{vth\_e}$ in the ASIC configuration.
Increasing these parameters by one digital-to-analog converter (DAC) step is equal to increasing the trigger level by approximately \SI{2.5}{\milli\volt}, \SI{15}{\milli\volt}, and \SI{20}{\milli\volt}, respectively, over a baseline set during calibration \cite{PETsys_priv_comm}.
For a proper operation of the trigger logic \cite{schug2018TOFPET2}, it has to be ensured that the voltage thresholds at the discriminators $\mathsf{Vth\_T1}$ and $\mathsf{Vth\_T2}$ fulfill $\mathsf{Vth\_T1} < \mathsf{Vth\_T2}$.
The trigger circuit of each channel enables dark count rejection and high timing resolution by triggering on a low voltage threshold with the first discriminator $\mathsf{D\_T1}$ at a very early point in time, i.e., on the first optical photons hitting the SiPM.
This trigger is delayed by a specified delay period and passes an AND gate opened by a second trigger activated on a higher voltage threshold with a second discriminator $\mathsf{D\_T2}$ (see Fig. \ref{fig:schematic_circuit}).
The delay period is in the order of few nanoseconds and can be configured by adjusting the parameter $\mathsf{fe\_delay}$ of the ASIC configuration \cite{PETsysTOFPET2DataSheet}.
The design of the trigger logic allows the rejection of small noise pulses, but is associated with the generation of satellite peaks in the coincidence time difference spectra which are caused by a shift in timestamp generation from the delayed $\mathsf{D\_T1}$ to the non-delayed $\mathsf{D\_T2}$.
A detailed description on the operation of the trigger circuit and the generation of satellite peaks is given in \cite{schug2018TOFPET2}.

\begin{figure}[h!]
	\includegraphics[width=0.45\textwidth]{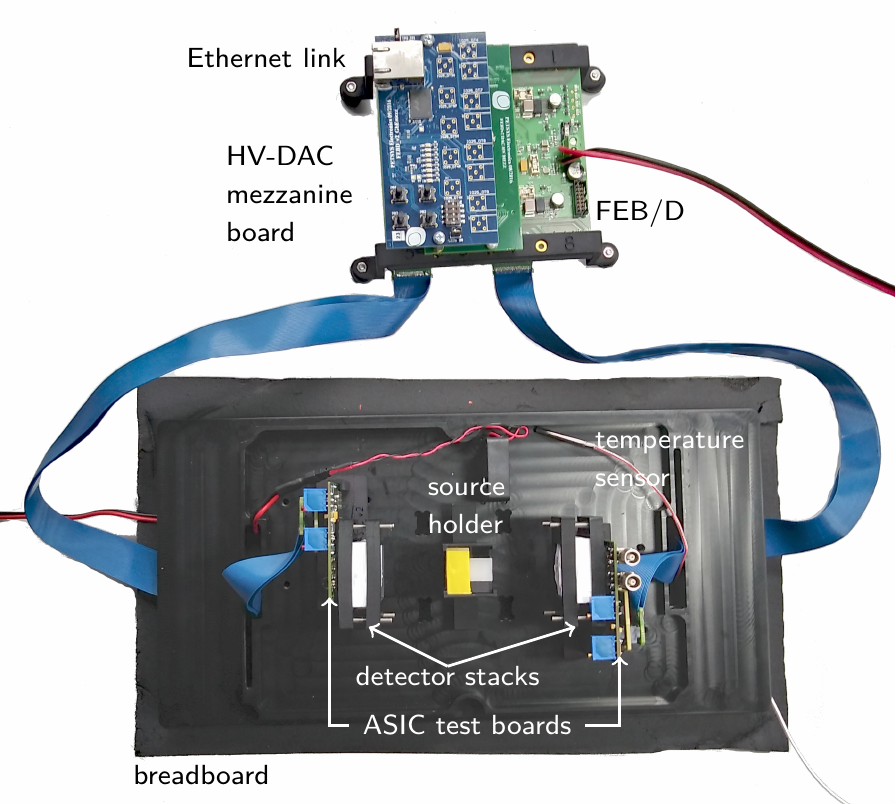}
	\caption{\csentence{Benchtop setup.}
		Optical breadboard holding two ASIC test boards equipped with KETEK PA3325-WB-0808 SiPMs for coincidence experiments. The whole setup can be enclosed with a top cover featuring thermo-control and operated inside a climate chamber. Graphic reprinted from \cite{nadig2019TOFPET2PowerConsumption}.}
	\label{fig:setup}
\end{figure}

\subsection{Benchtop setup}

\begin{figure}[h!]
	\includegraphics[width=0.45\textwidth]{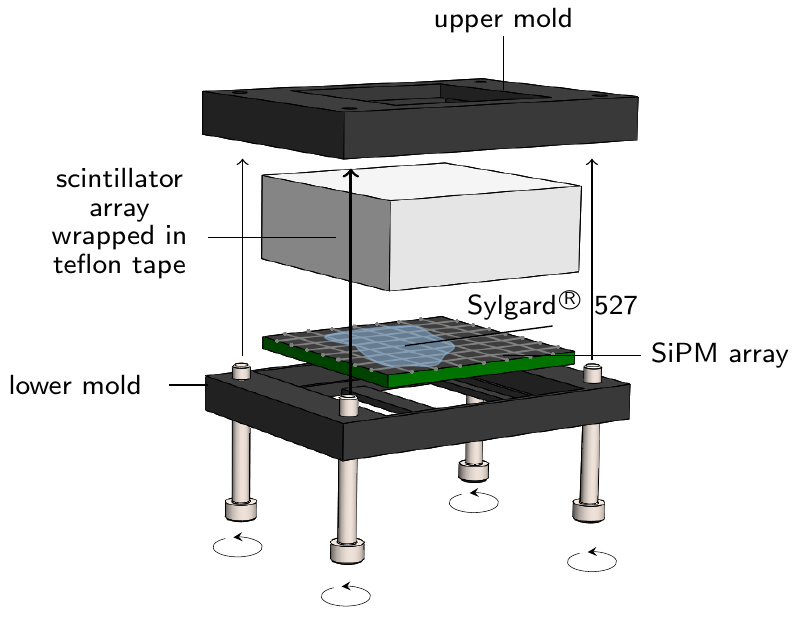}
	\caption{\csentence{Assembly tools.}
		Molds used to assemble the detector stacks composed of a segmented scintillator array, a coupling material, and an SiPM array. Each stack was wrapped in teflon tape to prevent light-loss.}
	\label{fig:gluing_tool}
\end{figure}

To evaluate the TOFPET2 ASIC in combination with different SiPM types, we used the TOFPET2 ASIC evaluation kit designed by PETsys Electronics S.A. as it enables the user to test various ASIC-SiPM combinations under benchtop conditions.
The kit comes along with two ASIC test boards each allowing to connect an SiPM array to a TOFPET2 ASIC via two SAMTEC connectors.
A high-voltage digital-to-analog converter (HV-DAC) mezzanine board (version 08/2016) providing the bias voltage for the employed SiPMs, and a front end board (FEB/D) generating the global clock signal and holding the main power supply for the ASIC test boards are included as well. 
Data transmission to a readout computer is conducted via a 1-Gigabit Ethernet (GbE) mezzanine board.
The ASIC test boards provide direct coupling between the employed SiPMs and the ASIC.\\
Additionally, a breadboard for mounting the setup is provided with the kit (see Fig. \ref{fig:setup}).
To conduct coincidence experiments, the breadboard allows to position both ASIC test boards face-to-face in different distances to the source holder mounting position.
Featuring two cable inlets, which were used to connect the ASIC test boards to the FEB/D board via two flexible cables, the breadboard can be encapsulated light-proof by a top cover.
A temperature sensor and a proportional–integral–derivative (PID) controller regulating a thermo-element were used to adjust the temperature insight the box.
In addition, the whole benchtop setup was placed in a climate chamber to ensure a stable temperature control.
A custom-designed source holder was employed to assemble the multi-source geometry.
Molds were used to assemble the crystal-SiPM array configurations (see Fig. \ref{fig:gluing_tool}).

\subsection{Detectors}

To test the compatibility and performance of the TOFPET2 ASIC with different analog SiPM types, we used 8 $\times$ 8 SiPM arrays fabricated by different vendors.
Two samples each of KETEK PA3325-WB-0808, SensL ArrayJ-30020-64P-PCB, Hamamatsu S14161-3050-HS-08, and Broadcom AFBR-S4N44P643S (see Tab. \ref{tab:sipm_characteristics}) were used to perform coincidence experiments.
A multi-source geometry of five \textsuperscript{22}Na NEMA cube point sources with a total activity of approximately \SI{3}{\mega\becquerel} was placed in the center of the setup during these experiments.
For the SensL ArrayJ-30020-64P-PCB SiPM arrays, additional adapter boards needed to be employed to establish the connection to the ASIC test board.\\
Using Sylgard\textsuperscript{\textregistered} 527, a two-component dielectric gel fabricated by Dow Corning for optical coupling, each SiPM array was successively one-to-one-coupled to 8 $\times$ 8 or 4 $\times$ 4 LYSO scintillator arrays. These arrays feature different segmentation layers (\SI{360}{\micro\m} or \SI{110}{\micro\m} BaSO$_4$ powder mixed with epoxy and \SI{67}{\micro\m} glued enhanced specular reflector (ESR) foil.
The size of an individual scintillator needle in the 8 $\times$ 8 ESR array is \SI{3.28x3.28x12.00}{\milli\m}.
The size of an individual scintillator needle in the 4 $\times$ 4 ESR array is \SI{3.85x3.85x12.00}{\milli\m}.
For thicker reflector layers, the scintillator width is reduced by the additional width of the reflector.
The selected scintillator length was chosen to reduce the influence of time jitter related light-transport effects accompanying long scintillator needles \cite{lecoq2017PushingTheLimits}, and thus, be more sensitive to ASIC-related effects on the timing performance. 
The chosen scintillator length can be found in some pre-clinical and clinical PET systems \cite{vandenberghe2016RecentDevelopments,schug2016petperformance}.
A performance degradation is expected if operating the TOFPET2 ASIC in combination with longer scintillator needles.
To give a benchmark of this degradation, additional measurements were conducted using a 4 $\times$ 4 ESR array consisting of \SI{2.62x2.62x19.00}{\milli\m} scintillator needles.

\subsection{Setup calibration}

The setup was calibrated using the PETsys calibration routine implemented in the software coming along with the evaluation kit \cite{TOFPET2EKitSoftwareGuide}. 
For each SiPM type, a calibration was run once at default ASIC configuration applying an overvoltage of \SI{4}{\volt} at an environment temperature of \SI{16}{\celsius}.
The required bias voltages were specified depending on the SiPM type employed.
It is not necessary to calibrate the setup for every threshold or overvoltage change.
A radioactive source can be left in the setup during calibration since this does not affect the determined parameters \cite{PETsys_priv_comm}.

\subsection{Parameter studies}

For each SiPM-crystal configuration, the discriminator threshold $\mathsf{vth\_t1}$ was varied between 10 and 50 in steps of 10, while $\mathsf{vth\_t2}$ and $\mathsf{vth\_e}$ were kept at constant values ($\mathsf{vth\_t2} = 20$, $\mathsf{vth\_e} = 15$).
This roughly corresponds to triggering between the \SIrange{1}{3}{\photoelectron} (photoelectron) as it is reported for the example of a Hamamatsu S14161-3050-HS-08 SiPM array in \cite{nadig2019TOFPET2PowerConsumption}.
For each setting, raw data were acquired for \SI{120}{\second} at an overvoltage of \SIrange{2.75}{7.75}{\volt}.
Additionally, the crystal top-to-source distance was varied between \SIrange{18}{58}{\mm} in steps of \SI{20}{\mm}.\\
For the KETEK PA3325-WB-0808 arrays coupled to the scintillator arrays featuring \SI{360}{\micro\m} BaSO$_4$ as segmentation layer, the trigger delay period was configured as different lengths in the range of few nanoseconds (\SIrange{0.0}{12.9}{\nano\second}) for $\mathsf{vth\_t1} = 10$ and $\mathsf{vth\_t1} = 50$ and overvoltages between \SIrange{2.75}{7.75}{\volt}.
For the Hamamatsu S14161-3050-HS-08 arrays coupled to the scintillator arrays featuring \SI{360}{\micro\m} BaSO$_4$ as segmentation layer, the integrator gain settings $\mathsf{G_{Q1}}$ and $\mathsf{G_{Q2}}$ were varied between 0.32-2.25 and 1.00-1.68, respectively, to evaluate the linearity of the acquired energy value spectra and the influence on the resulting energy resolution.
For each setting, raw data were acquired for \SI{120}{\second} at an overvoltage of \SI{4.75}{\volt} and with $\mathsf{vth\_t1} = 20$, $\mathsf{vth\_t2} = 20$, and $\mathsf{vth\_e} = 15$.
Effects due to the assembly of SiPM and scintillator array can be excluded since groups of measurements varying the trigger delay and gain configuration as well as the discriminator thresholds are performed with the same assembly directly after each other.

\DIFaddbegin 

\DIFaddend \subsection{Data collection and processing}

For data collection, we used the data acquisition routine implemented by PETsys.
We prepared the acquired raw data with the $\mathsf{convert\_raw\_to\_singles}$ method implemented by PETsys and provided with the evaluation kit \cite{TOFPET2EKitSoftwareGuide}.
Using this routine, raw data were converted into single-raw-hit information by applying the acquired calibration data.
A table containing a timestamp, an energy value, and a channel id for each single event acquired was returned.
The returned energy value does not correspond to ADC bins on hardware level, but is given in an arbitrary unit.\\
These single events were further processed using an analysis software developed at our institute \cite{schug2018TOFPET2}.
The energy values returned by the PETsys routine were sorted into channel-individual energy value histograms.
Afterwards, the noise background of this spectrum was estimated and subtracted \cite{Burgess1983431,Morhac1997113,Ryan1988396}.
The positions of the \SI{511}{\keV} and \SI{1274.5}{\keV} peaks were determined applying a Gaussian peak finder routine and an iterative Gaussian fitting method after subtracting the background of the spectrum \cite{Morhac2000108}.
A simple saturation model neglecting offsets and optical crosstalk was applied to the found positions, parametrizing the saturation corrected energy as
\begin{equation}
E = c \cdot s \cdot \log \left( \frac{1}{1 - \frac{e}{s}} \right)
\end{equation}
where $E$ is the energy in keV, $c$ is a correction factor in calibrated energy units$^{-1}$, $s$ is a saturation factor in calibrated energy units and $e$ is the acquired energy value in calibrated energy units \cite{schug2018TOFPET2}, similar to \cite{bugalho2017ExperimentalResultsTOFPET2}.
Now, all acquired energy values are converted and sorted into channel-individual energy histograms.
An energy filter ranging from \SIrange{400}{700}{\keV} was applied to filter true coincidences, which should have an energy around \SI{511}{\keV}, from scattered events.
A Gaussian was fitted to the remaining histogram data corresponding to the \SI{511}{\keV} peak.
The energy resolution was determined as the full width at half maximum (FWHM) of the fitted Gaussian summing up all channel-individual energy histograms to a global energy histogram.
The fit range was iteratively adjusted by \SI{10}{\percent} of the FWHM.
Filtered events were checked for coincidences applying a coincidence window of \SI{7.5}{\nano\second}.
To evaluate the effects of different trigger delay configurations, this window was extended to \SI{35}{\nano\second}.
The time difference between two events matched as a coincidence was calculated.
All computed time differences were filled into a time difference histogram.
A Gaussian was fitted to the histogram peak while iteratively adjusting the fit range by \SI{10}{\percent} of the FWHM.
The coincidence resolution time (CRT) was determined as the FWHM of the fitted Gaussian.
Additionally, the full width at tenth maximum (FWTM) is computed to get information about the tail behavior of the peak.
A Gaussian excess factor $\mathsf{gaussexc}$ is used to specify the outgrow of the tails with reference to their expected behavior.
This factor is defined as \cite{schug2017timewalk}
\begin{equation}
	\mathsf{gaussexc} = \frac{\mathsf{FWTM} - \frac{\mathsf{FWHM}}{2.355} \cdot 4.294}{\mathsf{FWTM}}
\end{equation}
where $\mathsf{FWHM}/2.355 \cdot 4.294$ computes the theoretical value of the FWTM for a truly Gaussian distribution.
The obtained performance parameters were always plotted as a function of the offset-corrected overvoltage $U_{\mathsf{ov\_cor}}$ computed via
\begin{equation}
U_{\mathsf{ov\_cor}} = U_{\mathsf{ov,set}} - U_{\mathsf{off}} 
\end{equation}
The overvoltage voltage set $U_{\mathsf{ov,set}}$ via the acquisition routine was corrected by an offset $U_{\mathsf{off}}$ (approximately \SI{750}{\milli\volt} \cite{TOFPET2EKitSoftwareGuide2019}).

\section{Results}

\begin{figure*}[h!]
	\includegraphics[width=0.95\textwidth]{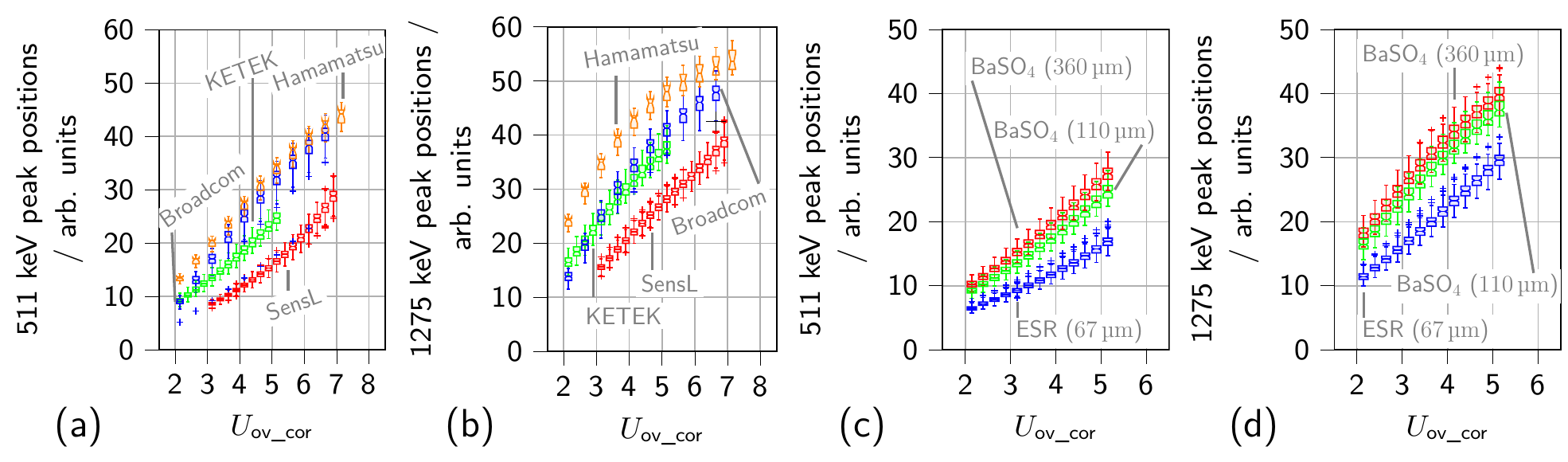}
	\caption{\csentence{Spread in gain for 128 ASIC channels coupled to SiPM arrays.}
		A geometry of five NEMA cubes (\textsuperscript{22}Na sources) with a total activity of approximately \SI{3}{\mega\becquerel} is employed.  (a) Position of the 511-keV peak in the energy value spectrum for different SiPM types. (b) Position of the 1275-keV peak in the energy value spectrum. Measurements were conducted using \SI{110}{\micro\m} BaSO$_4$ as segmentation layer and SiPM arrays from Broadcom, KETEK, SensL, and Hamamatsu. (c) Position of the 511-keV peak in the energy value spectrum for different segmentation layers. (d) Position of the 1275-keV peak in the energy value spectrum for different segmentation layers. Measurements were conducted with two KETEK PA3325-WB-0808 SiPM arrays. All positions are plotted against the applied offset-corrected overvoltage $\mathsf{U_{ov\_cor}}$.}
	\label{fig:channel_gain}
\end{figure*}

\begin{figure*}[h!]
	\includegraphics[width=0.95\textwidth]{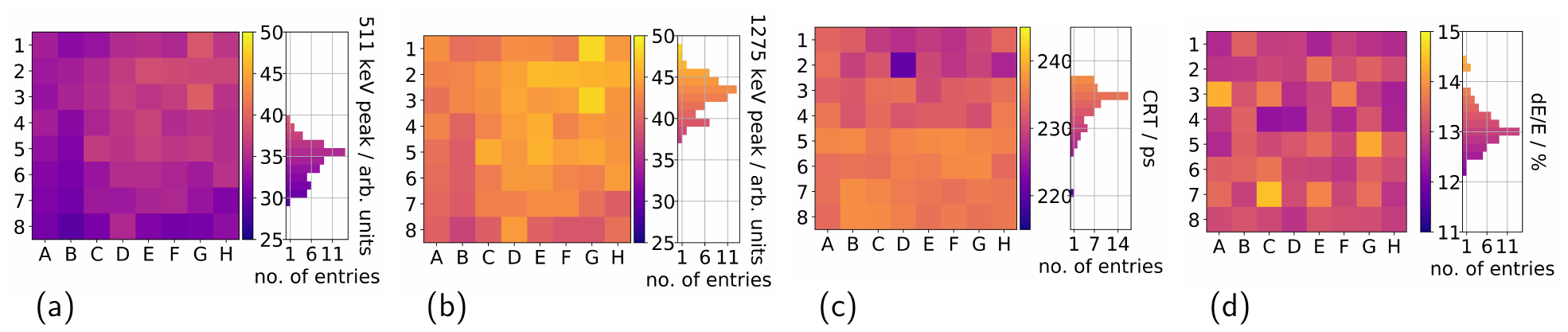}
	\caption{\csentence{Uniform channel performance.}
		A geometry of five NEMA cubes (\textsuperscript{22}Na sources) with a total activity of approximately \SI{3}{\mega\becquerel} is employed.  (a) Position of the 511-keV peak in the energy value spectrum. (b) Position of the 1275-keV peak in the energy value. (c) Coincidence resolution time (CRT) computed as the mean of the CRTs of this channel with all its coincident channels. (d) Energy resolution (dE/E).  Measurements were conducted with two Hamamatsu S14161-350-HS-08 SiPM arrays at \SI{4}{\volt} overvoltage and $\mathsf{vth\_t1} = 50$. BaSO$_4$ powder (\SI{360}{\micro\m}) mixed with an epoxy was used as segmentation layer. The scintillator array used consists of 8 $\times$ 8 needles with a size of \SI{3x3x12}{\milli\m}.}
	\label{fig:colormaps}
\end{figure*}

\begin{figure*}[h!]
	\includegraphics[width=0.9\textwidth]{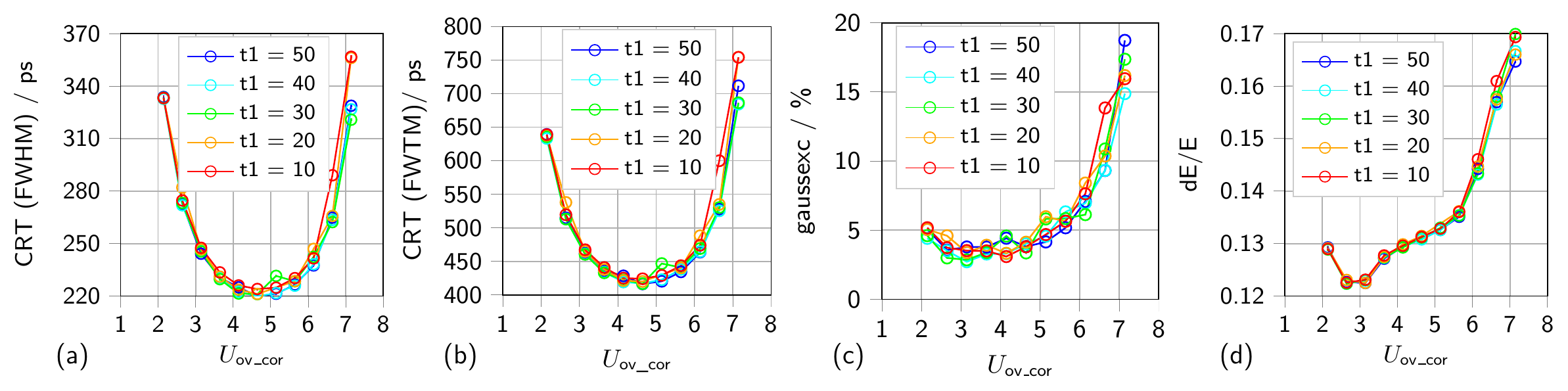}
	\caption{\csentence{Impact of the trigger threshold on CRT and energy resolution.}
		Measurements were conducted with two Hamamatsu S14161-3050-HS-08 each coupled to an 8 $\times$ 8 12-mm-high LYSO scintillator array featuring \SI{360}{\micro\m} BaSO$_4$ powder mixed with epoxy as segmentation layer. Single scintillator needles had a size of \SI{3x3x12}{\milli\m}. A geometry of five NEMA cubes (\textsuperscript{22}Na sources) with a total activity of approximately \SI{3}{\mega\becquerel} is employed.  (a) Coincidence resolution time (CRT - FWHM). (b) Coincidence resolution time (CRT - FWTM). (c) Energy resolution (dE/E).}
	\label{fig:threshold_variation}
\end{figure*}

The energy value spectra of different SiPM types allow to determine 511-keV and 1275-keV peak positions for all operated channels.
The peaks are shifted to higher energy values for higher overvoltages (see Fig. \ref{fig:channel_gain}a and Fig. \ref{fig:channel_gain}b).
The spread of the positions observed for individual channels increases for higher overvoltages and the peak positions are sorted according to the SiPM gains %
(see Tab. \ref{tab:sipm_characteristics}).
In addition, the peaks are shifted to higher energy values if a thicker segmentation layer is used in the scintillator array and if BaSO$_4$ is used instead of ESR (see Fig. \ref{fig:channel_gain}c and Fig. \ref{fig:channel_gain}d).
The spatial structure of the channel spread of the peak positions as well as the determined energy resolution and CRT is shown exemplary for a Hamamatsu S14161-3050-HS-08 array in Fig. \ref{fig:colormaps}.
The channel spread in peak position and energy resolution shows a Gaussian behavior with few outliers.
For the CRTs, which were calculated as mean of the CRTs of the respective channel with its coincident channels, a longer tail of the distribution is visible towards lower values.\\
Determining these performance parameters globally, operating two Hamamatsu S14161-3050-HS-08 SiPM arrays at a range of overvoltages shows an improvement in CRT (FWHM and FWTM) and energy resolution for higher overvoltages before deteriorating again (see Fig. \ref{fig:threshold_variation}). 
The reported Gaussian excess factor $\mathsf{gaussexc}$ show an increased contribution of non-Gaussian tails to the time difference spectra for higher overvoltages (see Fig. \ref{fig:threshold_variation}c).

\begin{figure*}[h!]
	\includegraphics[width=0.95\textwidth]{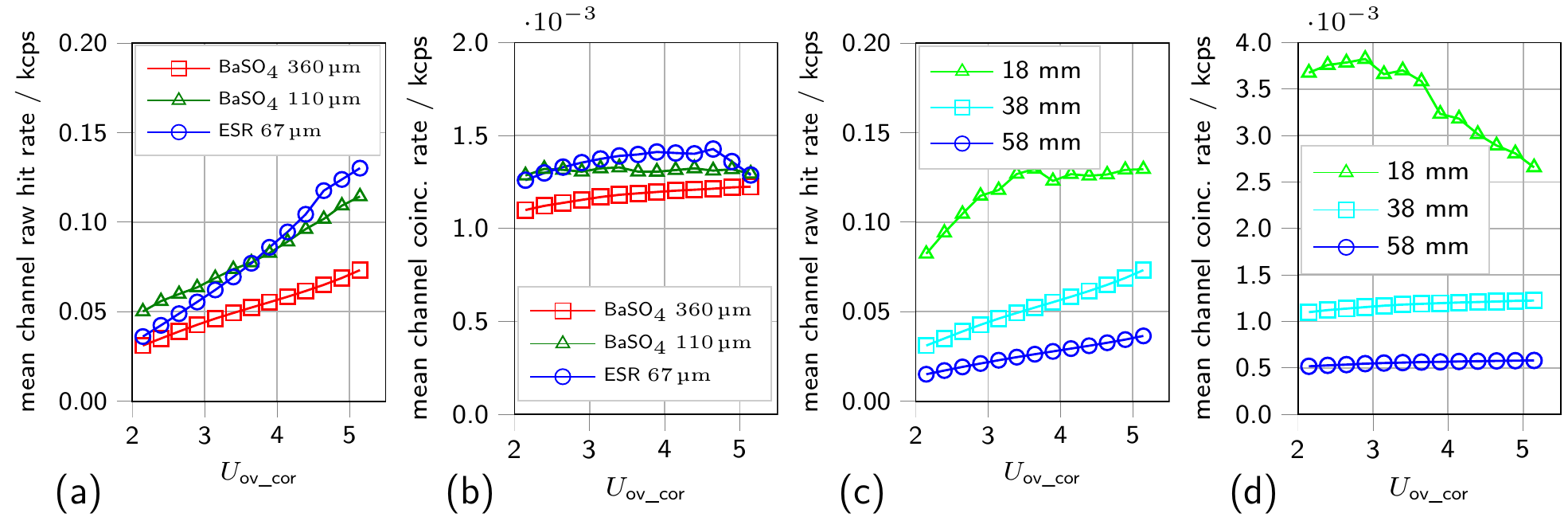}
	\caption{\csentence{Data rates for different setup hardware configurations.}
		(a) Mean raw hit rate per channel for different reflector materials. (b) Mean coincidence rate per channel for different reflector materials. (c) Mean raw hit rate per channel for different distances of the detectors to the source. (d) Mean coincidence rate per channel for different distances of the detectors to the source. }
	\label{fig:datarates}
\end{figure*}

Triggering on higher thresholds $\mathsf{vth\_t1}$ further improves the CRT (FWHM and FWTM). 
The energy resolution is rarely affected by this parameter change.
Generally, these effects are observed for all investigated SiPM types and segmentation layers.  
The point of transition from the negative to the positive influence of higher overvoltages and discriminator thresholds $\mathsf{vth\_t1}$ varies according to the SiPM type and scintillator array used.\\
To compare the influence of the investigated parameters, the lowest CRT (FWHM) achieved with each configuration along the investigated range of overvoltages is reported together with the corresponding energy resolution in Tab. \ref{tab:multi_channel_parameter_study_CRT} and Tab. \ref{tab:multi_channel_parameter_study_eres}.
Depending on the SiPM type, changing the threshold $\mathsf{vth\_t1}$ results in a CRT (FWHM) improvement of approximately \SIrange{3}{9}{\pico\second} (approximately \SIrange{1}{3}{\percent} relative improvement).
Triggering on a higher threshold, also leads to slightly lower Gaussian excess factors, i.e., a more Gaussian behavior of the tails of the coincidence time difference histogram, especially for KETEK and SensL SiPMs (see Tab. \ref{tab:multi_channel_parameter_study_CRT_FWTM}).
Among the different scintillator arrays, both scintillator types employing BaSO$_4$ powder mixed with epoxy as segmentation layer outperform the type featuring ESR foil.
A segmentation layer of \SI{360}{\micro\m} BaSO$_4$ reaches approximately \SIrange{100}{200}{\pico\second} lower CRTs (FWHM) than a segmentation layer of \SI{67}{\micro\m} ESR foil (see Tab. \ref{tab:multi_channel_parameter_study_CRT}).
This is equal to a relative performance gain of approximately \SIrange{29}{43}{\percent} depending on the SiPM type.
The CRT (FWTM) shows a similar behavior (see Tab. \ref{tab:multi_channel_parameter_study_CRT_FWTM}).
Comparing the corresponding energy resolutions, an absolute improvement of up to approximately \SI{2.5}{\percent} can be reported (see Tab. \ref{tab:multi_channel_parameter_study_eres}).
Here, the thicker BaSO$_4$-layer of \SI{360}{\micro\m} shows better performance results (approximately \SIrange{6}{14}{\percent} relative performance gain) than the thinner layer of \SI{110}{\micro\m}.
For different segmentation layers, different mean channel raw hit rates and  mean channel coincidence rates, respectively, are reported (see Fig. \ref{fig:datarates}a and \ref{fig:datarates}b).\\
Triggering at larger distances between sensor and the employed multi-source geometry leads to an improvement of the CRT (FWHM) and energy resolution by about \SIrange{10}{20}{\pico\second} and \SI{1.7}{\percent} for a distance variation of \SI{20}{\mm} (see Tab. \ref{tab:multi_channel_parameter_study_CRT} and \ref{tab:multi_channel_parameter_study_eres}).
Moving the detector stacks from the farthest to the closest distance corresponds to a three- to four-fold increase of the mean channel raw hit rate and an about seven-fold increase of the mean channel coincidence rate, measured at \SI{2}{\volt} (see Fig. \ref{fig:datarates}c and \ref{fig:datarates}d).\\
Due to limited availability of materials, the parameter study was only partly performed for the Broadcom AFBR-S4N44P643S SiPM arrays.
Here, triggering on a higher threshold $\mathsf{vth\_t1}$ improved the lowest CRT (FWHM) achieved by up to approximately \SI{25}{\pico\second}, which corresponds to approximately \SI{9}{\percent} relative performance gain (see Tab. \ref{tab:multi_channel_parameter_study_Broadcom}).
The corresponding CRT (FWTM) is improved as well.
The corresponding energy resolution only showed minor fluctuations less than \SI{0.3}{\percent} (absolute change).
Again, the scintillator type employing BaSO$_4$ powder mixed with epoxy as segmentation layer outperform the type featuring ESR foil regarding the CRT.\\ %
Among the four SiPM types tested in combination with the TOFPET2 ASIC, the 8 $\times$ 8 Hamamatsu S14161-3050-HS-08 SiPM arrays outperform the other SiPM arrays achieving CRTs (FWHM) down to \SI{219.9}{} $\pm$ \SI{0.7}{\pico\second} (dE/E = \SI{13.08}{} $\pm$ \SI{0.03}{\percent}) with an 8 $\times$ 8 12-mm-high scintillator array featuring \SI{360}{\micro\m} BaSO$_4$ mixed with epoxy as segmentation layer for 128 channels read out.
Broadcom AFBR-S4N44P643S SiPM arrays reached comparable CRTs down to \SI{216.1}{} $\pm$ \SI{2.6}{\pico\second} (dE/E = \SI{9.46}{} $\pm$ \SI{0.09}{\percent}) with a 4 $\times$ 4 12-mm-high scintillator array featuring a thinner BaSO$_4$ layer (\SI{110}{\micro\m}) for 32 channels read out.
With a 4 $\times$ 4 19-mm-high scintillator array featuring \SI{155}{\micro\m} air-coupled ESR-foil as the inter-crystal layer, the Hamamatsu arrays reached CRTs (FWHM) down to \SI{247.5}{} $\pm$ \SI{2.4}{\pico\second} (dE/E = \SI{10.46}{} $\pm$ \SI{0.08}{\percent}) for 32 channels read out.\\

\begin{figure*}[h!]
	\includegraphics[width=0.95\textwidth]{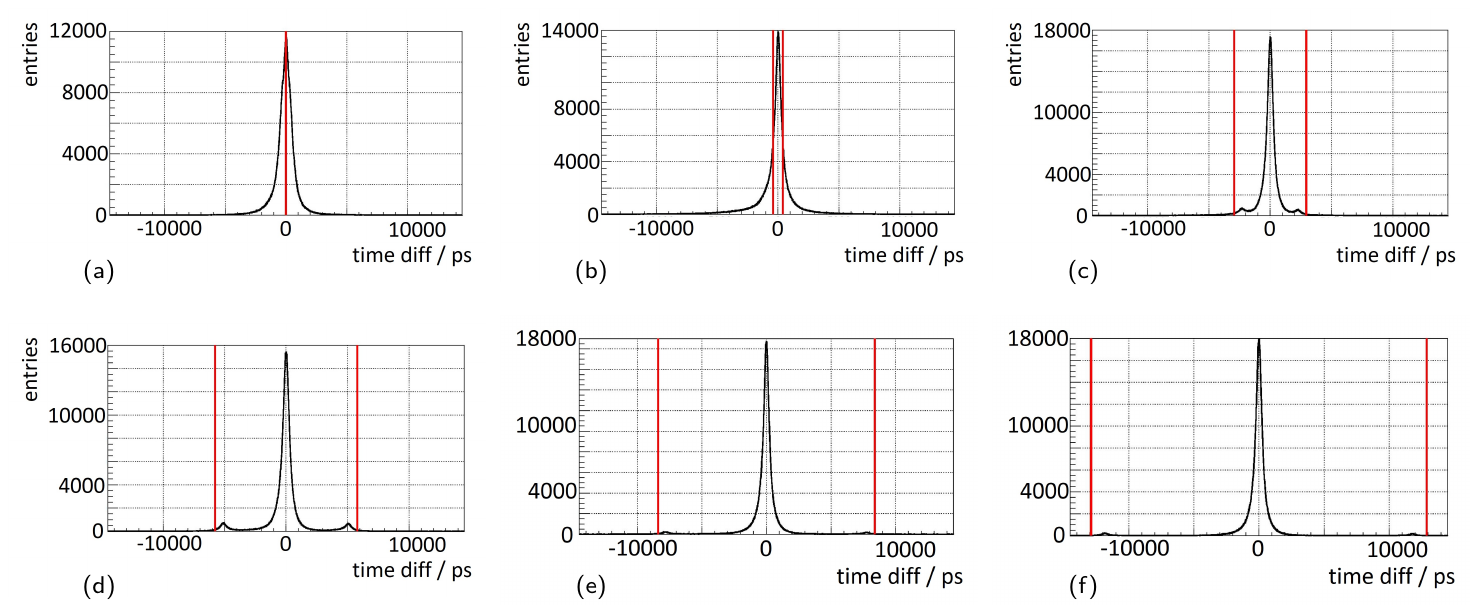}
	\caption{\csentence{Satellite peaks in the coincidence time difference spectra for different delay periods.}
		Measurements were conducted with two KETEK PA3325-WB-0808 each coupled to an 8 $\times$ 8 12-mm-high LYSO scintillator array featuring \SI{360}{\micro\m} BaSO$_4$ powder mixed with epoxy as segmentation layer.
		Data are collected at \SI{4.75}{\volt} overvoltage and with $\mathsf{vth\_t1} = 10$. A geometry of five NEMA cubes (\textsuperscript{22}Na sources) with a total activity of approximately \SI{3}{\mega\becquerel} is employed. Red lines indicate the configured delay period \cite{PETsys_priv_comm,PETsysTOFPET2DataSheet}. (a) Delay line bypassed. (b) \SI{0.39}{\nano\second}. (c) \SI{2.95}{\nano\second}. (d) \SI{5.8}{\nano\second}. (e) \SI{8.4}{\nano\second}. (f) \SI{12.9}{\nano\second}.}
	\label{fig:fe_delay_configs}
\end{figure*}

\begin{figure*}[h!]
	\includegraphics[width=0.95\textwidth]{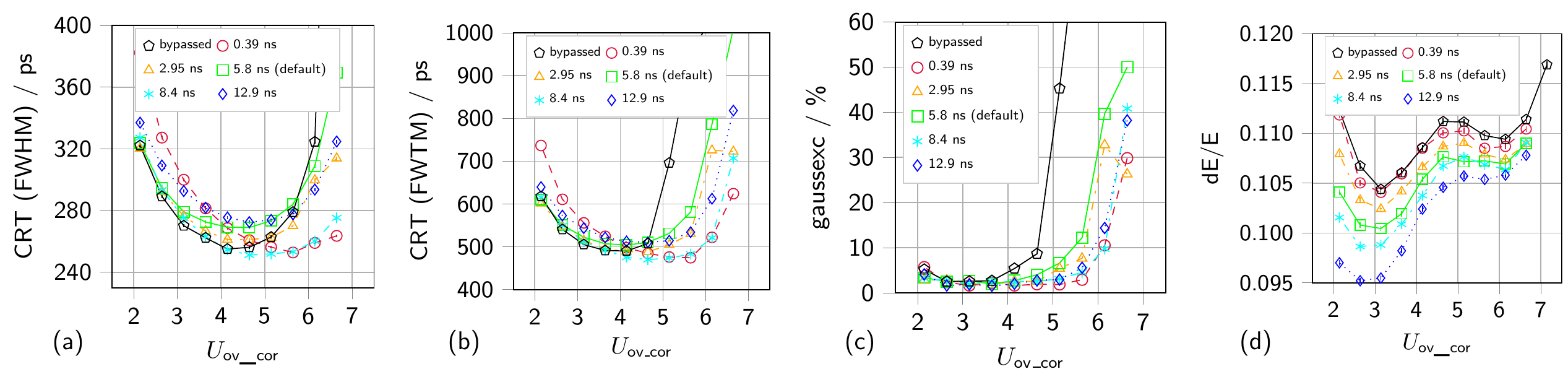}
	\caption{\csentence{Performance results dependent on the configured delay period.}
		Data are acquired with two KETEK PA3325-WB-0808 in multi-channel coincidence experiments with $\mathsf{vth\_t1} = 50$. Each SiPM array is coupled to an 8 $\times$ 8 LYSO of \SI{12}{\mm} height scintillator array featuring \SI{360}{\micro\m} BaSO$_4$ powder mixed with epoxy as the inter-crystal layer. A geometry of five NEMA cubes (\textsuperscript{22}Na sources) with a total activity of approximately \SI{3}{\mega\becquerel} is employed. (a) Coincidence resolution time (CRT - FWHM). (b) Coincidence resolution time (CRT - FWTM). (c) Energy resolution (dE/E).}
	\label{fig:fe_delay}
\end{figure*}

Satellite peaks, which were observed for measurements with single SiPMs \cite{schug2018TOFPET2}, also appear in the coincidence time difference spectra of multi-channel measurements (see Fig. \ref{fig:fe_delay_configs}).
The peaks are generated by small noise pulses that trigger the first discriminator and are validated by a true coincidence event triggering on the second discriminator occurring shortly after the noise event and while the $\mathsf{AND}$ gate connecting first and second discriminator is still active.
As Fig. \ref{fig:fe_delay_configs} depicts, the shift of the peaks from the center peak in the time difference histogram is related to the configured trigger delay between the first and second discriminator of the ASIC channel circuit.
The fraction of events causing the formation of satellite peaks is higher for higher overvoltages and lower discriminator thresholds $\mathsf{vth\_t1}$.
A broader explanation is given in \cite{schug2018TOFPET2}.
As Fig. \ref{fig:fe_delay} depicts, changing the delay period between the first and second discriminator, not only leads to the shift of satellite peaks over the range of the coincidence time difference spectrum but also causes a systematic deterioration in energy resolution for shorter delay periods (relative deterioration of about \SI{10}{\percent}).
Additionally, the lowest CRTs (FWHM) achieved vary by up to approximately \SI{30}{\pico\second} (up to approximately \SI{11}{\percent} relative change). %
The lowest FWTM of the coincidence time difference histograms varies by up to approximately \SI{50}{\pico\second}.
For higher overvoltages, the effect of a narrower peaks and less distinct tails in the time difference spectra depending on the delay configurations reverses itself (see Fig. \ref{fig:fe_delay}b).
The computed Gaussian excess factors show that especially for higher overvoltage the influence of non-Gaussian histogram tails become more and more evident (see Fig. \ref{fig:fe_delay}c).
Repeating the measurements and evaluation with different discriminator thresholds $\mathsf{vth\_t1}$ or narrower coincidence windows did not lead to a change of these effects.\\

\begin{figure*}[h!]
	\includegraphics[width=0.95\textwidth]{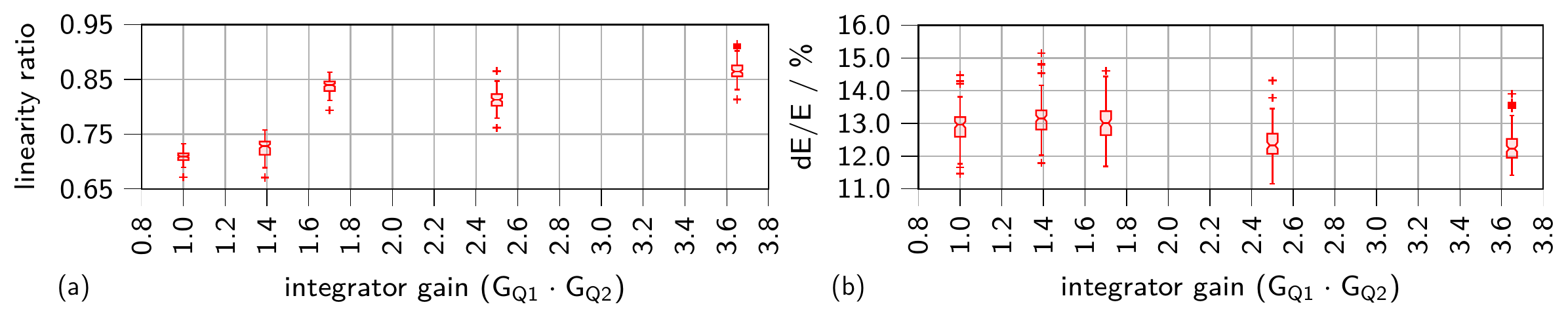}
	\caption{\csentence{Energy linearity and energy resolution for different integrator gain configurations.}
		The default configuration is $\mathsf{G_{Q1}} \cdot \mathsf{G_{Q2}} = 1.0$. (a) Energy linearity computed as the ratio of the \SI{511}{\keV} and \SI{1274.5}{\keV} peak positions in the energy value histogram. (b) Energy resolution (dE/E) computed using the FWHM of the \SI{511}{\keV} peak in the energy histogram.}
	\label{fig:integgain}
\end{figure*}

Fig. \ref{fig:integgain} depicts the linearity ratio computed as the ratio of the \SI{511}{\keV} and \SI{1275}{\keV} peak positions in the ADC value spectra as well as the energy resolution for different integrator gains $\mathsf{G_ {Q1}}$ and $\mathsf{G_{Q2}}$.
We find that the default integrator gain setting $\mathsf{G_ {Q1} = 1.0}$ and $\mathsf{G_{Q2} = 1.0}$ results in the most linear energy spectra (linearity ratio of 0.71, dE/E $\approx$ \SI{13}{\percent}).
The energy resolution can be improved by approximately \SI{5}{\percent} (relative improvement) changing the $\mathsf{G_{Q1}}$-setting to $\mathsf{G_ {Q1} = 2.5}$ at the cost of energy linearity. %
The absolute spread in energy resolution for both of these settings is on the order of approximately \SI{0.6}{\percent}.
The spread is generally increased by higher gains, which also result in a deterioration of the energy resolution to up to \SI{14}{\percent}.

\section{Discussion}

The peak positions reported for the energy value spectra (see Fig. \ref{fig:channel_gain} and Fig. \ref{fig:colormaps}) suggest that all channels on one ASIC operate uniformly. 
Since especially in Fig. \ref{fig:channel_gain} 128 channel, i.e., two ASICs, are considered when computing the channel spread, the operation of two different ASICs is uniform as well. 
Since single channels not matching the peak position pattern were only identified for one of the Hamamatsu S14161-3050-S-08 SiPM arrays (outliers in Fig. \ref{fig:channel_gain}a and \ref{fig:channel_gain}b), but not for other SiPM arrays, this behavior is probably due to the connected SiPMs and not the operation of ASIC channels. 
This channel was excluded from further evaluation steps.
Since all channels operate uniformly, reporting the CRT and energy resolution as globally computed parameters is justified.\\
In single- and multi-channel studies, a similar behavior of the CRT and energy resolution dependent on the applied overvoltage was observed.
With the selected thresholds, a convergence of the CRT (FWHM) can already be observed for SensL and Hamamatsu SiPMs when setting vth\_t1=40 and vth\_t1=50 (see Tab. \ref{tab:multi_channel_parameter_study_CRT}). 
The improvement observed for KETEK SiPMs is less than \SI{2}{\pico\second}.
For Broadcom SiPMs, a slight convergence is visible yet, as the improvement becomes less for higher thresholds (see Tab. \ref{tab:multi_channel_parameter_study_Broadcom}). 
Additionally, one can observe a slight decrease in energy resolution for higher thresholds (see Tab. \ref{tab:multi_channel_parameter_study_eres} and Tab. \ref{tab:multi_channel_parameter_study_Broadcom}).
This can be explained by the fact that only higher pulses, i.e., events with a higher energy, are triggered on with higher thresholds, leading to more compressed energy spectra.
Since for most cases, a convergence of the CRT and a deterioration of the energy resolution is observed, we refrained from triggering on higher thresholds.
The non-Gaussian tails of the peak in the coincidence time difference histogram remain below extension \SI{10}{\percent} of the Gaussian peak at the optimal operation point of the respective SiPM and are suspected to be caused by the detection of scattered events. 
This matches the observation that the tails are more prone for low thresholds $\mathsf{vth\_t1}$ and higher overvoltages, increasing the gain of the SiPM, making it more sensitive to noise and low-energy events.
Thus, we suspect that the tails could be reduced by applying a narrower energy window. 
This effect of the energy window on the CRT (FWTM) was investigated for digital SiPMs in \cite{schug2017timewalk}.
The impact of the discriminator threshold $\mathsf{vth\_t1}$ on the achieved performance is stronger in multi-channel than in previous single-channel investigations \cite{schug2018TOFPET2}.
While in single-channel measurements with two KETEK PM3325-WB-A0 the effect was only visible for overvoltages lower than \SI{3}{\volt} and no reverse effect was observed for higher overvoltages, for multi-channel measurements with two KETEK PA3325-WB-0808, the impact of $\mathsf{vth\_t1}$ was visible over the whole investigated overvoltage range and reversed itself for overvoltages higher than \SI{4}{\volt}.
The observed differences in behavior could be due to effects on the SiPM array such as crosstalk and light-sharing.\\
Among the different parameters changed, the scintillator segmentation layer has the largest relative impact on the achieved performance (\SIrange{29}{43}{\percent}).
The effect of improved performance for BaSO$_4$ as reflector material is likely due to a lower probability of light-sharing between channels and thus, a higher fraction of $\gamma$-events depositing their whole energy on one SiPM channel.
This leads to a higher number of small pulses being acquired resulting in a higher mean raw hit rate and especially mean coincidence rate per channel.
The observed higher raw data rate with increasing overvoltage is expected due to the increase of the SiPM gain, crosstalk and photo-detection efficiency with the overvoltage (see Fig. \ref{fig:datarates}a).
The coincidence rate is expected and observed to be stable over the investigated overvoltage range (see Fig. \ref{fig:datarates}b).
Since no saturation is observed, the performance dependency on the reflector material is probably not a $\gamma$-rate effect.
In addition, higher energy values are observed for individual channels if thicker segmentation layers are used as shown in Fig. \ref{fig:channel_gain}c and Fig. \ref{fig:channel_gain}d.
The different reflective behaviors of BaSO$_4$ (diffusive reflection) and ESR foil (specular reflexion) also suspected to contribute to the reduced and increased event rate and more or less light-sharing between single scintillator needles.
Setting higher thresholds $\mathsf{vth\_t2}$ and $\mathsf{vth\_e}$ could be used to filter out these events, while still triggering on the first optical photons with a low thresholds $\mathsf{vth\_t1}$, which preserve the ToF information of the detected event.
An improved performance seen in experiments with larger distances to the employed sources, which also is associated with a lower mean raw hit rate per channel and lower mean coincidence rate per channel (see Fig. \ref{fig:datarates}c and Fig. \ref{fig:datarates}d), additionally indicates a raw-data-rate-dependent performance.
Here, a saturation of the mean raw hit rate per channel is visible for the detector position closest to the sources at higher overvoltage (see Fig. \ref{fig:datarates}c).
The detected coincidence rate drops accordingly (see Fig. \ref{fig:datarates}d).
Since here, the detector configuration stayed the same and only the distance to the source was varied, this is a $\gamma$-rate effect.
Since the origin of the loss of events is not flagged, it cannot be determined if this drop is due to the ASIC or other parts of the signal chain. 
PETsys Electronics S.A. states a maximum event rate of up to \SI{480}{kcps} per channel \cite{PETsysTOFPET2DataSheet}, which is not reached by the detected mean raw hit rate per channel.\\
Communication with PETsys lead to the assumption that the systematic deterioration in energy resolution with changing the trigger delay period is probably related to the configuration of the energy integration window.
Acquiring data in qdc-mode, the integrated charge is corrected by an estimate for a charge offset related to the time difference between $\mathsf{trigger\_Q}$, which starts the energy integration, and the delayed $\mathsf{trigger\_T1}$, which sets the event timestamp, i.e., the time of arrival of the optical photons hitting the SiPM channel.
The latter one is prolonged or shortened with respect to the configuration of the trigger delay period, which results in an adaptation of the charge offset applied as a correction.
The energy integration window is not adapted for different trigger delay periods and thus, needs to be adjusted manually. 
In all investigations so far, this window has been kept at a fixed default value of approximately \SI{290}{\nano\second}.
Further experiments to investigate the influence of manually adjusting the window are on-going.
The difference in behavior as a function of overvoltage cannot be understood without profound knowledge of the details of the ASIC implementation.
Discussions with PETsys regarding this matter are on-going.\\
Investigations with two Hamamatsu S14161-3050-HS-08 arrays show that the integrator gain setting can be used to linearize the acquired energy spectra and optimize the energy resolution.
The absolute spread in the energy resolution of all investigated channels is in the order of less than \SI{1}{\percent}.
Thus, it can be concluded that the SiPM and ASIC channels operated uniformly in this experiment.
In this study, the ASIC is already operated at the most linear setting possible.
This setting is recommended to be used for further investigations featuring high-gain SiPMs.
Switching to a more non-linear setting to improve the energy resolution at \SI{511}{\keV} is possible as long as the saturation correction method can be applied, i.e., as long as the two peaks of the Na\textsuperscript{22}-spectrum remain separable.\\
Comparing the performance of different analog SiPM types, Hamamatsu S14161-3050-HS-08 and Broadcom AFBR-S4N44P643S show the lowest CRTs in combination with the TOFPET2 ASIC. 
Both SiPM types feature a higher gain and a larger SPAD size than the investigated KETEK and SensL SiPMs (see Tab. \ref{tab:sipm_characteristics}).
However, it has to be kept in mind that this study only reports performance results for the combination of SiPM type and ASIC using the delivered benchtop setup "as-is".
Since the coupling between SiPM and ASIC was not individually optimize for each SiPM type, this study cannot be used to compare the performance of solely the SiPM types among each other, but only their performance in combination with the TOFPET2 ASIC for the specific coupling scheme given by the electronics provided.
Individual optimization of the SiPM-ASIC coupling is expected to improve the reported performance \cite{huizenga2012FastPreamp,acerbi2019SiPMSimulation}.\\
On benchtop level, the TOFPET2 ASIC measures up to the performance of state-of-the-art ASICs in single-channel coincidence experiments. 
An overview on single-channel experiments is provided in \cite{schug2018TOFPET2}. 
In comparison with other multi-channel performance experiments, the TOFPET2 ASIC measures up to state-of-the-art system performance regarding its ToF capability.
An 18-channel ASIC using a tot-method to digitize energies achieved a CRT of \SI{275}{\pico\second} and an energy resolution of \SI{11.8}{\percent} in coincidence experiments with two 12 $\times$ 12 Hamamatsu C13500-4075LC-12 modules coupled to 20-mm-high scintillator arrays and read out by 8 ASICs in total \cite{goertzen2019HamamatsuCRT}.
Module tests with the PETA5 ASIC and SiPMs fabricated by Fondazione Bruno Kessler (FBK, RGB-HD technology) achieved and average CRT of \SI{230}{\pico\second} and an energy resolution of about \SI{14}{\percent} when one-to-one-coupled to 10-mm-high LYSO scintillator arrays \cite{sacco2015PETA5}.
PETsys reports a CRT of \SI{260}{\pico\second} between two Hamamatsu S13361-3050AE-04 arrays each one-to-one-coupled to an array of 15-mm-high LYSO needles \cite{bugalho2017ExperimentalResultsTOFPET2}. 
On system level, the TRIMAGE scanner equipped with the TRIROC ASIC reaches CRTs of \SI{515}{\pico\second} and an energy resolution between \SIrange{20}{22}{\percent} employing a two-layered scintillator geometry consisting of LYSO needles with a total height of \SI{20}{\mm} on FBK SiPMs (NUV-HD technology) \cite{belcari2019TRIMAGE}.
A study on a prototype PET scanner employing digital SiPMs (DPC 3200-22 by Philips Digital Photon Counting) coupled to 10-mm-high LYSO scintillator arrays reports CRTs down to \SI{215.2}{} $\pm$ \SI{0.4}{\pico\second} (energy resolution: \SI{11.381}{} $\pm$ \SI{0.007}{\percent}) \cite{schug2015tofring}.
The Biograph Vision PET/CT (computed tomography) scanner recently released by Siemens also achieves \SI{214}{\pico\second} on system level, using \SI{3.2x3.2x20}{\mm} LSO needles \cite{BiographVisionDataSheet,BiographWebPage}.
Further performance studies should include monolithic scintillator blocks as well as slabs to obtain DOI information with the TOFPET2 ASIC.
First experiments featuring DOI-capable scintillator geometries have been conducted \cite{peng2019ComptonPET,li2019DOI,lamprou2020TOFPET2Monoliths}.\\
In addition, when designing a PET system or a MR-compatible PET insert, one has to keep in mind other system requirements such as power supply limitations or interference problems \cite{vandenberghe2015PETMRIchallenges}.
The power consumption of the TOFPET2 ASIC and its influence on the ASIC performance were already evaluated in an additional study \cite{nadig2019TOFPET2PowerConsumption}.
Regarding MR-compatibility tests, similar test protocols as in \cite{weissler2015MRtests,wehner2015MRcompatibilityassessment,wehner2014InvestigationOfMRcompatibility} need to be defined and executed.

\section{Outlook and Conclusion}

Generally, the TOFPET2 ASIC evaluation kit enables the user to easily test various SiPM types and scintillator topologies in combination with the TOFPET2 ASIC.\\
The ASIC is compatible with all SiPM types tested in this study  (see Tab. \ref{tab:sipm_characteristics}).
We observe that high discriminator thresholds ($\mathsf{vth\_t1} = 40 - 50$), thicker reflector layers, and lower data rates (larger distances to the sources and diffusive inter-crystal reflexion) result in an improved performance regarding both CRT and energy resolution.
Among the SiPM types used, so far, Broadcom AFBR-S4N44P643S and Hamamatsu S14161-3050-HS-08 SiPMs show the most promising performance results in combination with the TOFEPT2 ASIC.\\
The appearance of satellite peaks is confirmed for multi-channel time difference spectra. 
A performance dependency related to the configuration of the delay element is observed.
A first indication of a $\gamma$-rate-dependent ASIC performance was found in experiments with different distances to a Na\textsuperscript{22} source.
Adjustments of the integrator gain can be used to linearize the energy value spectra acquired in qdc-mode or to improve the energy resolution.\\
Even without optimizing the SiPM-to-ASIC coupling, the CRTs and energy resolutions reported for a coincidence setup of two detector blocks lay in the order of the performance of clinical and pre-clinical ToF-PET systems, while providing a uniform and stable readout of multiple channels at the same time. 
Therefore, it will be considered for integration in (whole-body) ToF-PET/MRI applications.\\

\DIFaddbegin

\DIFaddend \begin{backmatter}

\section{Abbreviations}

ASIC -- application-specific integrated circuit\\
BGO -- bismuth germanate\\
CRT -- coincidence resolution time\\
CT -- computed tomography
DOI -- depth of interaction\\
FBK -- Fondazione Bruno Kessler\\
FEB -- front end board\\
FWHM -- full width at half maximum\\
FWTM -- full width at tenth maximum\\
GbE -- Gigabit Ethernet\\
HV-DAC -- high-voltage digital-to-analog converter\\
LOR -- line of response\\
LYSO -- lutetium-yttrium oxyorthosilicate\\
MRI -- magnetic resonance imaging\\
PDE -- photo detection efficiency
PET -- positron emission tomography\\
QDC -- charge-to-digital converter\\
SiPM -- silicon-photomultiplier\\
TDC -- time-to-digital converter\\
ToF -- time-of-flight

\section*{Declarations}

\subsection*{Ethics approval and consent to participate}
Not applicable.

\subsection*{Consent for publication}
Not applicable.

\subsection*{Availability of data and material}
The datasets used and/or analyzed during the current study are available from the corresponding author on reasonable request.

\subsection*{Competing interests}
The authors declare that they have no competing interests.

\subsection*{Funding}
This project has received funding from the European Union's Horizon 2020 research and innovation programme under grant agreement No 667211 \protect\includegraphics[height=0.75em]{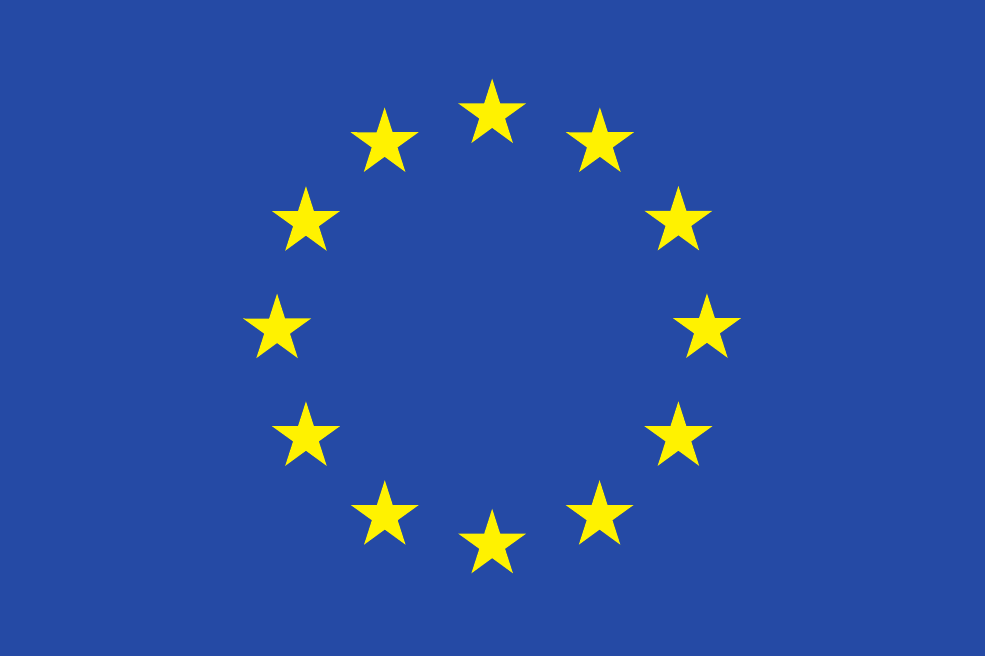}.

\subsection*{Authors' contributions}
All authors participated in the design of the study.
VN contributed the experiments and data evaluation and wrote the manuscript. 
DS established a framework for data analysis.
BW designed additional electronic parts for the setup. 
All authors contributed to the manuscript by discussing and revising its content. 
All authors gave their approval for the final version of the manuscript.

\subsection*{Acknowledgments}

We thank Ricardo Bugalho and Luis Ferramacho from PETsys Electronics S.A. for kindly answering our many questions.
We thank Torsten Solf from Philips Digital Photon Counting for sharing his expertise and for providing additional supplies.

\bibliographystyle{bmc-mathphys} %
\bibliography{bibliography}      %

\section*{Tables}

\begin{table*}[h!]
	\caption{Characteristics of the investigated analog SiPM types. Parameters are taken from the respective data sheet.}
	\label{tab:sipm_characteristics}
	\begin{tabular}{p{2cm}p{2cm}p{2.2cm}p{2cm}p{2.2cm}}
		\hline
		& KETEK & SensL & Hamamatsu & Broadcom\\
		& PA3325-WB-0808 & Array-J-30020-64P-PCB &  S14161-3050-HS-08 & AFBR-S4N44P643S \\
		\hline
		Pitch & \SI{3.36}{\mm} & \SI{3.36}{\mm} & \SI{3.20}{\mm} & \SI{3.93}{\mm} \\
		Active area & \SI{3.0}{} $\times$ \SI{3.0}{\mm\squared} & \SI{3.07}{} $\times$ \SI{3.07}{\mm\squared} & \SI{3.0}{} $\times$ \SI{3.0}{\mm\squared} & \SI{3.72}{} $\times$ \SI{3.72}{\mm\squared} \\
		No. of SPADs & 13920 & 14850 & 3531 & 15060\\
		SPAD size & \SI{25}{\micro\m} & \SI{20}{\micro\m} & \SI{50}{\micro\m} (pitch) & \SI{30}{\micro\m} (pitch)\\
		Breakdown voltage & \SI{27}{\volt} & \SIrange{24.2}{24.7}{\volt} & \SI{37}{\volt} & \SI{26.9}{\volt} \\
		Gain & $1.74 \cdot 10^6$ (\SI{5}{\volt}) & $1.9 \cdot 10^6$ (\SI{5}{\volt}) & $2.5 \cdot 10^6$ (\SI{2.7}{\volt}) & $3.3 \cdot 10^6$ (\SI{7}{\volt}) \\
		Dark count rate & \SI{100}{\kilo\hertz\per\mm\squared} (\SI{5}{\volt}) & \SI{125}{\kilo\hertz\per\mm\squared} (\SI{5}{\volt}) & - & \SI{270}{\kilo\hertz\per\mm\squared} (\SI{7}{\volt})\\
		PDE & \SI{43}{\percent} (\SI{5}{\volt}) & \SI{38}{\percent} (\SI{5}{\volt}) & \SI{50}{\percent} (\SI{2.7}{\volt}) & \SI{55}{\percent} (\SI{7}{\volt})\\
		Reference & \cite{KETEKProductDataSheetArray} & \cite{SensLProductDataSheetSingle} & \cite{HamamatsuProductDataSheetArray} & \cite{BroadcomProductDataSheetArray} \\
		\hline
	\end{tabular}
\end{table*}

\begin{table*}[h!]
	\caption{Coincidence resolution time (FWHM). Parameter study for multi-channel configurations featuring SiPMs by different vendors. Table reports the lowest CRT achieved in coincidence measurements of two times 64 channels with the corresponding settings and materials used. If not indicated differently, data were acquired at $\mathsf{vth\_t1 = 50}$, $\mathsf{vth\_t2 = 20}$, $\mathsf{vth\_e = 15}$, with \SI{360}{\micro\m} BaSO$_4$ for scintillator segmentation,  a scintillator height of \SI{12}{\milli\m}, and at a distance of \SI{38}{mm}. The overvoltage setting varies according to the parameter setting and SiPM type used. For the clinical configuration (19-mm high scintillator), only two times 16 channels were read out.}
	\label{tab:multi_channel_parameter_study_CRT}
	\centering
	\begin{tabular}{lccc}
		\hline
		& KETEK & SensL & Hamamatsu\\
		& PA3325-WB-0808 & Array-J-30020-64P-PCB & S14161-3050-HS-08\\	
		\hline	
		& & &  \\
		\hspace{0mm} threshold $\mathsf{vth\_t1}$ & & &  \\
		\hspace{3mm} 10 & $273.8 \pm 0.7$ \SI{}{\ps} & $265.2 \pm 0.6$ \SI{}{\ps} & $224.1 \pm 0.7$ \SI{}{\ps}  \\
		\hspace{3mm} 20 & $271.4 \pm 0.7$ \SI{}{\ps} & $265.4 \pm 0.6$ \SI{}{\ps} & $221.2 \pm 0.7$ \SI{}{\ps}  \\
		\hspace{3mm} 30 & $268.6 \pm 0.7$ \SI{}{\ps} & $263.7 \pm 0.6$ \SI{}{\ps} & $221.1 \pm 0.7$ \SI{}{\ps}  \\
		\hspace{3mm} 40 & $266.1 \pm 0.6$ \SI{}{\ps} & $262.8 \pm 0.6$ \SI{}{\ps} & $219.9 \pm 0.7$ \SI{}{\ps}  \\
		\hspace{3mm} 50 & $264.6 \pm 0.6$ \SI{}{\ps} & $262.4 \pm 0.6$ \SI{}{\ps} & $220.0 \pm 0.7$ \SI{}{\ps}  \\
		& & & \\
		\hspace{0mm} segmentation layer & & & \\
		\hspace{3mm} BaSO$_4$ (\SI{360}{\micro\m}) & $264.6 \pm 0.6$ \SI{}{\ps} & $262.4 \pm 0.6$ \SI{}{\ps} & $220.0 \pm 0.7$ \SI{}{\ps} \\
		\hspace{3mm} BaSO$_4$ (\SI{110}{\micro\m}) & $282.8 \pm 0.6$ \SI{}{\ps} & $306.5 \pm 0.6$ \SI{}{\ps} & $242.9 \pm 0.5$ \SI{}{\ps} \\
		\hspace{3mm} ESR (\SI{67}{\micro\m}, glued) & $371.4 \pm 0.8$ \SI{}{\ps} & $463.4 \pm 0.8$ \SI{}{\ps} & $328.1 \pm 0.7$ \SI{}{\ps} \\
		& & & \\
		\hspace{0mm} distance & & &  \\
		\hspace{3mm} \SI{18}{\mm} & $277.3 \pm 0.4$ \SI{}{\ps} & - & - \\
		\hspace{3mm} \SI{38}{\mm} & $264.6 \pm 0.6$ \SI{}{\ps} & $262.4 \pm 0.6$ \SI{}{\ps} & $220.0 \pm 0.7$ \SI{}{\ps}  \\
		\hspace{3mm} \SI{58}{\mm} & $245.3 \pm 0.8$ \SI{}{\ps} & $253.3 \pm 0.8$ \SI{}{\ps} & $224.4 \pm 0.7$ \SI{}{\ps}  \\
		& & & \\
		\hspace{0mm} clinical (\SI{19}{\milli\m} height) & & &  \\
		\hspace{3mm} ESR (\SI{155}{\micro\m}, air-coupled) & $317.3 \pm 1.9$ \SI{}{\ps} & $325.1 \pm 1.7$ \SI{}{\ps} & $247.5 \pm 2.4$ \SI{}{\ps}  \\
		\hline
	\end{tabular}
\end{table*}  

\begin{table*}[h!]
	\caption{Coincidence resolution time (FWTM) and Gaussian excess factors $\mathsf{gaussexc}$. Parameter study for multi-channel configurations featuring SiPMs by different vendors. Table reports the CRT (FWTM) and Gaussian excess factors corresponding to the lowest CRTs achieved in coincidence measurements of two times 64 channels (see Tab. \ref{tab:multi_channel_parameter_study_CRT}). If not indicated differently, data were acquired at $\mathsf{vth\_t1 = 50}$, $\mathsf{vth\_t2 = 20}$, $\mathsf{vth\_e = 15}$, with \SI{360}{\micro\m} BaSO$_4$ for scintillator segmentation, a scintillator height of \SI{12}{\milli\m}, and at a distance of \SI{38}{mm}. The overvoltage setting varies according to the parameter setting and SiPM type used. The error on the reported CRTs (FWTM) is less than \SI{0.1}{\pico\second}. For the clinical configuration (19-mm high scintillator), only two times 16 channels were read out.}
	\label{tab:multi_channel_parameter_study_CRT_FWTM}
	\centering
	\begin{tabular}{lcccccc}
		\hline
		& \multicolumn{2}{c}{KETEK} & \multicolumn{2}{c}{SensL} & \multicolumn{2}{c}{Hamamatsu}\\
		& \multicolumn{2}{c}{PA3325-WB-0808} & \multicolumn{2}{c}{Array-J-30020-64P-PCB} & \multicolumn{2}{c}{S14161-3050-HS-08}\\	
		& & & & & & \\
		& CRT (FWTM) & gaussexc & CRT (FWTM) & gaussexc & CRT (FWTM) & gaussexc  \\
		\hline	
		& & & & & & \\
		\hspace{0mm} threshold $\mathsf{vth\_t1}$ & & & & & & \\
		\hspace{3mm} 10 & $524.6$ \SI{}{\ps} & $5.07$ \SI{}{\percent} & $528.9$ \SI{}{\ps} & $9.38$ \SI{}{\percent} & $424.1$ \SI{}{\ps} & $3.80$ \SI{}{\percent} \\
		\hspace{3mm} 20 & $514.2$ \SI{}{\ps} & $3.89$ \SI{}{\percent} & $514.6$ \SI{}{\ps} & $6.34$ \SI{}{\percent} & $420.1$ \SI{}{\ps} & $4.13$ \SI{}{\percent}  \\
		\hspace{3mm} 30 & $511.4$ \SI{}{\ps} & $4.40$ \SI{}{\percent} & $517.2$ \SI{}{\ps} & $7.56$ \SI{}{\percent} & $416.7$ \SI{}{\ps} & $3.36$ \SI{}{\percent} \\
		\hspace{3mm} 40 & $501.6$ \SI{}{\ps} & $3.40$ \SI{}{\percent} & $504.3$ \SI{}{\ps} & $5.22$ \SI{}{\percent} & $417.0$ \SI{}{\ps} & $4.02$ \SI{}{\percent} \\
		\hspace{3mm} 50 & $503.6$ \SI{}{\ps} & $4.39$ \SI{}{\percent} & $507.7$ \SI{}{\ps} & $6.12$ \SI{}{\percent} & $416.4$ \SI{}{\ps} & $3.82$ \SI{}{\percent} \\
		& & & & & & \\
		\hspace{0mm} segmentation layer & & & & & & \\
		\hspace{3mm} BaSO$_4$ (\SI{360}{\micro\m}) & $503.6$ \SI{}{\ps} & $4.39$ \SI{}{\percent} & $507.7$ \SI{}{\ps} & $6.12$ \SI{}{\percent} & $416.4$ \SI{}{\ps} & $3.82$ \SI{}{\percent} \\
		\hspace{3mm} BaSO$_4$ (\SI{110}{\micro\m}) & $542.6$ \SI{}{\ps} & $5.23$ \SI{}{\percent} & $604.9$ \SI{}{\ps} & $5.29$ \SI{}{\percent} & $464.9$ \SI{}{\ps} & $4.97$ \SI{}{\percent} \\
		\hspace{3mm} ESR (\SI{67}{\micro\m}, glued) & $715.1$ \SI{}{\ps} & $5.61$ \SI{}{\percent} & $896.7$ \SI{}{\ps} & $6.13$ \SI{}{\percent} & $609.6$ \SI{}{\ps} & $8.37$ \SI{}{\percent} \\
		& & & & & & \\
		\hspace{0mm} distance & & &  & & & \\
		\hspace{3mm} \SI{18}{\mm} & $539.6$ \SI{}{\ps} & $6.71$ \SI{}{\percent} & - & - & - & -\\
		\hspace{3mm} \SI{38}{\mm} & $503.6$ \SI{}{\ps} & $4.39$ \SI{}{\percent} & $507.7$ \SI{}{\ps} & $6.12$ \SI{}{\percent} & $416.4$ \SI{}{\ps} & $3.82$ \SI{}{\percent} \\
		\hspace{3mm} \SI{58}{\mm} & $463.8$ \SI{}{\ps} & $3.68$ \SI{}{\percent} & $477.2$ \SI{}{\ps} &  $3.32$ \SI{}{\percent}  & $419.9$ \SI{}{\ps} & $2.64$ \SI{}{\percent} \\
		& & & & & &  \\
		\hspace{0mm} clinical (\SI{19}{\milli\m} height) & & & & & & \\
		\hspace{3mm} ESR (\SI{155}{\micro\m}, air-coupled) & $601.9$ \SI{}{\ps} & $4.04$ \SI{}{\percent} & $617.2$ \SI{}{\ps} & $4.11$ \SI{}{\percent} & $462.3$ \SI{}{\ps} & $2.43$ \SI{}{\percent} \\
		\hline
	\end{tabular}
\end{table*}  

\begin{table*}[h!]
	\caption{Energy resolution.	Parameter study for multi-channel configurations featuring SiPMs by different vendors. Table reports the energy resolution corresponding to the lowest CRTs achieved in coincidence measurements of two times 64 channels (see Tab. \ref{tab:multi_channel_parameter_study_CRT}). If not indicated differently, data were acquired at $\mathsf{vth\_t1 = 50}$, $\mathsf{vth\_t2 = 20}$, $\mathsf{vth\_e = 15}$, with \SI{360}{\micro\m} BaSO$_4$ for scintillator segmentation, a scintillator height of \SI{12}{\milli\m}, and at a distance of \SI{38}{mm}. The overvoltage setting varies according to the parameter setting and SiPM type used. For the clinical configuration (19-mm high scintillator), only two times 16 channels were read out.}
	\label{tab:multi_channel_parameter_study_eres}
	\centering
	\begin{tabular}{lccc}
		\hline
		& KETEK & SensL & Hamamatsu\\
		& PA3325-WB-0808 & Array-J-30020-64P-PCB & S14161-3050-HS-08\\		
		\hline	
		& & & \\
		\hspace{0mm} threshold $\mathsf{vth\_t1}$ & & & \\
		\hspace{3mm} 10 & $10.35 \pm 0.02$ \SI{}{\percent} & $11.20 \pm 0.02$ \SI{}{\percent} & $13.10 \pm 0.02$ \SI{}{\percent} \\
		\hspace{3mm} 20 & $10.44 \pm 0.02$ \SI{}{\percent} & $10.92 \pm 0.02$ \SI{}{\percent} & $13.15 \pm 0.03$ \SI{}{\percent} \\
		\hspace{3mm} 30 & $10.84 \pm 0.02$ \SI{}{\percent} & $11.22 \pm 0.02$ \SI{}{\percent} & $13.12 \pm 0.03$ \SI{}{\percent}\\
		\hspace{3mm} 40 & $10.85 \pm 0.02$ \SI{}{\percent} & $11.06 \pm 0.02$ \SI{}{\percent} & $13.08 \pm 0.03$ \SI{}{\percent} \\
		\hspace{3mm} 50 & $11.08 \pm 0.02$ \SI{}{\percent} & $11.23 \pm 0.02$ \SI{}{\percent} & $13.13 \pm 0.03$ \SI{}{\percent} \\
		& & & \\
		\hspace{0mm} segmentation layer & & & \\
		\hspace{3mm} BaSO$_4$ (\SI{360}{\micro\m}) & $11.08 \pm 0.02$ \SI{}{\percent} & $11.23 \pm 0.02$ \SI{}{\percent} & $13.13 \pm 0.03$ \SI{}{\percent} \\
		\hspace{3mm} BaSO$_4$ (\SI{110}{\micro\m}) & $12.60 \pm 0.02$ \SI{}{\percent} & $11.64 \pm 0.02$ \SI{}{\percent} & $14.70 \pm 0.02$ \SI{}{\percent} \\
		\hspace{3mm} ESR (\SI{67}{\micro\m}, glued) & $12.58 \pm 0.02$ \SI{}{\percent} & $13.78 \pm 0.02$ \SI{}{\percent} & $14.74 \pm 0.02$ \SI{}{\percent} \\
		& & & \\
		\hspace{0mm} distance & & & \\
		\hspace{3mm} \SI{18}{\mm} & $10.94 \pm 0.01$ \SI{}{\percent} & - & - \\
		\hspace{3mm} \SI{38}{\mm} & $11.08 \pm 0.02$ \SI{}{\percent} & $11.23 \pm 0.02$ \SI{}{\percent} & $13.13 \pm 0.03$ \SI{}{\percent} \\
		\hspace{3mm} \SI{58}{\mm} & $10.67 \pm 0.02$ \SI{}{\percent} & $10.66 \pm 0.02$ \SI{}{\percent} & $11.39 \pm 0.03$ \SI{}{\percent} \\
		& & & \\
		\hspace{0mm} clinical (\SI{19}{\milli\m} height) & & &  \\
		\hspace{3mm} ESR (\SI{155}{\micro\m}, air-coupled) & $10.74 \pm 0.05$ \SI{}{\percent} & $10.72 \pm 0.04$ \SI{}{\percent} & $10.46 \pm 0.08$ \SI{}{\percent} \\
		\hline
	\end{tabular}
\end{table*}  

\begin{table*}[h!]
	\caption{Supplementary results contributing to the parameter study. Table reports the lowest CRT achieved with the corresponding settings and materials used and the CRT (FWTM), Gaussian excess factors and energy resolution at the selected operation point. Data were acquired with two Broadcom AFBR-S4N44P643S SiPM array, where only 16 channels per array were read out.  If not indicated differently, data were acquired at $\mathsf{vth\_t1 = 50}$, $\mathsf{vth\_t2 = 20}$, $\mathsf{vth\_e = 15}$, with \SI{110}{\micro\m} BaSO$_4$ for scintillator segmentation,  a scintillator height of \SI{12}{\milli\m}, and at a distance of \SI{58}{mm}. The overvoltage setting varies according to the parameter setting used.}
	\label{tab:multi_channel_parameter_study_Broadcom}
	\centering
	\begin{tabular}{lcccc}
		\hline
		& \multicolumn{4}{c}{Broadcom AFBR-S4N44P643S}\\
		& & & & \\
		& CRT (FWHM) & CRT (FWHM) & gaussexc & dE/E \\	
		\hline	
		& & & & \\
		\hspace{0mm} threshold $\mathsf{vth\_t1}$ & & & & \\
		\hspace{3mm} 10 & $240.2 \pm 2.9$ \SI{}{\ps} & $449.7$ \SI{}{\ps} & $2.68$ \SI{}{\percent} & $9.37 \pm 0.08$ \SI{}{\percent} \\
		\hspace{3mm} 20 & $232.6 \pm 2.8$ \SI{}{\ps} & $437.4$ \SI{}{\ps} & $3.15$ \SI{}{\percent} & $9.38 \pm 0.08$ \SI{}{\percent} \\
		\hspace{3mm} 30 & $223.5 \pm 2.7$ \SI{}{\ps} & $426.9$ \SI{}{\ps} & $4.78$ \SI{}{\percent} & $9.65 \pm 0.09$ \SI{}{\percent} \\
		\hspace{3mm} 40 & $220.1 \pm 2.6$ \SI{}{\ps} & $419.1$ \SI{}{\ps} & $4.45$ \SI{}{\percent} & $9.40 \pm 0.08$ \SI{}{\percent} \\
		\hspace{3mm} 50 & $216.1 \pm 2.6$ \SI{}{\ps} & $413.3$ \SI{}{\ps} & $4.92$ \SI{}{\percent} & $9.46 \pm 0.09$ \SI{}{\percent} \\
		& & & & \\
		\hspace{0mm} segmentation layer & & & & \\
		\hspace{3mm} BaSO$_4$ (\SI{110}{\micro\m}) & $216.1 \pm 2.6$ \SI{}{\ps} & $582.0$ \SI{}{\ps} & $4.23$ \SI{}{\percent} & $9.46 \pm 0.09$ \SI{}{\percent} \\
		\hspace{3mm} ESR (\SI{67}{\micro\m}, glued) & $293.6 \pm 2.6$ \SI{}{\ps} & $413.3$ \SI{}{\ps} & $4.92$ \SI{}{\percent} & $11.64 \pm 0.08$ \SI{}{\percent}\\
		\hline
	\end{tabular}
\end{table*}  

\end{backmatter}
\end{document}